\documentclass[12pt,draftclsnofoot,onecolumn,english]{IEEEtran}
\usepackage[english]{babel}
\usepackage{cite}
\usepackage{amssymb, amsmath}
\usepackage{graphicx}
\usepackage{color}
\usepackage{url}
\usepackage{multirow}
\usepackage[ruled,vlined,linesnumbered]{algorithm2e}
\usepackage{subfigure}
\usepackage{dblfloatfix}

\definecolor{gold}{rgb}{0.85,.66,0}
\definecolor{cian}{rgb}{.02,.7,.95}
\definecolor{ppp}{rgb}{.06,.7,.55}

\definecolor{dblue}{rgb}{0,0,1} %

\usepackage{setspace}

\begin{document}
% Title
\title{{Quasi-Distributed} Antenna Selection for {Spectral Efficiency} Maximization in Subarray Switching  XL-MIMO Systems}

% Authors
%\author{\colm{João Henrique Inacio de Souza}, \colg{Abolfazl Amiri}, \colc{Taufik Abrão}, \colp{Elisabeth de Carvalho}, and \pp{Petar Popovski}
\author{
{João Henrique Inacio de Souza}, {Abolfazl Amiri}, {Taufik Abrão}, {Elisabeth de Carvalho}, and {Petar Popovski}

\thanks{J. H. I. de Souza and T. Abrão are with the Electrical Engineering Department, State University of Londrina, PR, Brazil.  E-mail: {joaohis@outlook.com; taufik@uel.br}}

\thanks{A.  Amiri, E. de Carvalho and P. Popovski are with the Department of Electronic Systems,  Technical Faculty of IT and Design; Aalborg University,	Denmark; E-mail: {aba@es.aau.dk; edc@es.aau.dk; petarp@es.aau.dk.}}
}

\maketitle

% Abstract
\begin{abstract}
In this paper, we consider the downlink (DL) of a zero-forcing (ZF) precoded extra-large scale massive MIMO (XL-MIMO) system. The base-station (BS) operates with limited number of radio-frequency (RF) transceivers due to high cost, power consumption and interconnection bandwidth associated to the fully digital implementation. The BS{, which is implemented with} a subarray switching architecture, selects groups of active antennas inside each subarray to transmit the DL signal. {This work proposes} efficient resource allocation (RA) procedures to perform joint antenna selection (AS) and power allocation (PA) to maximize the DL spectral efficiency (SE) of {an} XL-MIMO {system} operating under different {loading settings}. Two metaheuristic RA procedures based on the genetic algorithm (GA) are assessed and compared in terms of performance, coordination data size and computational complexity. {One algorithm is based on a quasi-distributed methodology while the other is based on the conventional centralized processing.} Numerical results demonstrate that the quasi-distributed GA-based procedure {results} in a suitable trade-off between performance, complexity and exchanged coordination data. {At the same time,} it {outperforms} the centralized procedures with appropriate system operation settings.
\end{abstract}

% Keyworks

% Introduction
\section{Introduction}\label{sec:introduction}
The benefits of adopting a high number of antennas at the base-station (BS) {have attracted} the interest on the massive MIMO transceiver design for the multi-antenna wireless communications systems beyond the fifth generation (B5G) and of the sixth generation (6G). The main advantages are the large array gain, inter-channel orthogonality and channel hardening. Also, increasing the number of antenna elements can {enhance} the cell coverage, improving the quality-of-service (QoS) of the border-cell users \cite{larsson2014}.

When the {BS array attains} extreme physical dimensions to support crowded scenario locations, such as airports and large shopping malls, the system is classified as extra-large scale massive MIMO (XL-MIMO) \cite{carvalho2019}. {The XL-MIMO array} provides the benefits of massive MIMO with additional beam-forming resolution due to the large array aperture \cite{martinez2014}. The XL-MIMO {array} is characterized by key changes in the electromagnetic propagation conditions {when compared to the conventional spatial stationary massive MIMO regime}. The first property is the spherical wavefront {propagation feature} for the received signal due to the distance between the BS and the users being less than the Rayleigh distance \cite{zhou2015}. Second, each cluster of scatterers sees only a portion of the array. Thus, the transmitted signal by each user {reaches a small} group of antennas, which {comprises} the visibility region {(VR)} of this user \cite{carvalho2019}. {Additionally}, the different propagation paths experienced along the array result in variations on the average received power. Results in \cite{li2015, ali2019} demonstrate that the {spatial} non-stationarities produced by these {two} properties {limit the performance of the system in terms of spectral efficiency (SE) unless an appropriated signal processing technique is applied.}
% can improve the spectral efficiency (SE) if appropriated signal processing is applied, and degrade the system performance otherwise.

% Channel model paragraph
%Recent works have proposed different channel models to account the XL-MIMO non-stationarities. In \cite{li2015} the non-stationarities are introduced as clusters of scatters partially visible to the array, in addition to the ones which are wholly visible. Differently, in \cite{amiri2018, marinello2020} the large-scale fading coefficient vary as the distance between the users and each antenna element in the array. Yet, in \cite{ali2019} the signal of each user is received by a small group of antennas. In \cite{amiri2020b}, is proposed an update of the double-scattering model to incorporate the non-stationarities, adding the concept of visibility region and the variation of the received power along the array.

Despite the benefits of high numbers of antennas, the XL-MIMO scenario imposes challenges for transceiver design. The first of them is the high cost and power consumption of fully digital implementations, which require one radio-frequency (RF) transceiver per antenna element \cite{garcia-rodriguez2017, gao2018}. In addition, adopting a large number of antennas demands a high interconnection bandwidth to transmit the baseband data throughout the links to the BS processing unit. This turns into a serious implementation bottleneck, since the required bandwidth can not be handled by the current radio interfaces \cite{li2017, sanchez2020}. Lastly, handling the complexity of signal processing techniques is a relevant issue, since the number of executed operations in linear detectors, such as zero-forcing (ZF) and minimum mean-squared error (MMSE), scales with the number of antennas \cite{mueller2016}.

In order to design practical {BS architectures}, one can limit the number of RF transceivers {to cope with the} cost constraints. The implementation {with a limited the number of RF transceivers} can benefit from the large array by adopting techniques {such} as antenna selection (AS) and hybrid precoding. Often, hybrid precoding design is associated with the solution of intricate optimization problems \cite{heath2016}. In addition, the commonly employed analog phase shifters are more expensive and consume more power than conventional on-off switches \cite{gao2018}. For these reasons, combining the AS procedures with linear precoding designs {result in attainable strategies aiming at} robust and effective implementations. Different approaches and tools can be adopted to perform AS, such as convex optimization \cite{dua2006, gao2015, garcia-rodriguez2017}, greedy heuristics \cite{garcia-rodriguez2017, lin2012}, machine learning \cite{amiri2020a} and metaheuristics \cite{marinello2020, lu2007, lain2011, makki2017}.

One strategy to combat the problem of high interconnection bandwidth is to use {hierarchical} architectures. Adding multiple processing units to handle small groups of antennas and choosing the right signal processing methods can reduce significantly the {amount} of exchanged information in the regime of asymptotic number of antennas, as discussed
in \cite{sanchez2020, li2017}. However, the coordination of such processing units to perform different signal processing and resource allocation (RA) tasks constitutes a big challenge. In addition, many of these activities rely on the knowledge of {fully reliable} channel state information (CSI), which is {hard to attain} due to the high array dimensions. Many works on channel estimation \cite{sanchez2019}, precoding and data detection \cite{amiri2018,amiri2019, amiri2020b, yang2019, sanchez2020, li2017} {in massive and XL-MIMO} consider distributed pre-processing at local nodes. However, {studies on the distributed RA strategies,} mainly involving AS, are scarce.

{The signal processing complexity is an important concern in XL-MIMO due to the high number of antenna elements. However, differently from the conventional massive MIMO, the XL-MIMO can benefit from the spatial non-stationarities adopting local signal processing strategies {to treat the signals inside the VRs at the BS' sub-arrays} with reduced complexity \cite{amiri2018,amiri2020b}.}

% Literature review
\subsection{Literature Review}\label{sec:literature-review}
{AS strategies for MIMO systems are extensively discussed in the literature. One AS algorithm to improve capacity in low rank matrix channels on point-to-point MIMO was first introduced in \cite{gore2000}. Later, the capacity distribution of systems with receive AS has been derived in \cite{molisch2005}. These results were extended to massive MIMO regime in \cite{asaad2018} and \cite{ouyang2019}. In these papers, the authors derived capacity bounds for systems with transmit and receive AS, respectively.}

The authors in \cite{dua2006, gao2015} proposed AS procedures {respectively for the channel capacity and downlink (DL) sum-capacity maximization} based on the convex optimization framework. One technique based on the branch-and-bound algorithm is {used} in \cite{gao2018}. Considering linearly-precoded systems, the problems of AS for SE and sum-SINR maximization are addressed respectively in \cite{lin2012, abdullah2020}. Differently, the work in \cite{siljak2018} analyzed one joint AS and power allocation (PA) procedure in a system with spatially distributed antennas. The proposed procedure runs at each antenna with side-information shared within its neighborhood. Besides, AS considering limited connections in the RF transceivers switching matrices is examined in \cite{garcia-rodriguez2017}.

{On the other hand,} there are only a few works that consider the AS problem for the XL-MIMO systems. A spatial users mapping procedure to maximize SE implemented with convolutional neural networks (CNN) {is proposed in \cite{amiri2020a}}. The aim is to determine each effective subarray window to precode the users signals using ZF. Results demonstrate that the CNN-based procedure achieves SE values comparable to the optimal mapping algorithm. In \cite{marinello2020}, several transmit AS procedures to maximize the energy efficiency (EE) from the long-term fading coefficients are proposed. Asymptotic SINR expressions for the received signal with AS are derived. Since the {derived} optimization problem is {NP-hard}, three of the proposed procedures are implemented by metaheuristic techniques, one being the genetic algorithm (GA). The GA is a powerful evolutionary metaheuristic that {was used} in different contexts to solve AS problems, as it is {considered} in \cite{lu2007, lain2011, makki2017}.

% Contribution
\subsection{Contribution}
Motivated by the benefits of large numbers of antennas at the BS and the restricted number of RF transceivers, this work examines the joint AS and PA problem on the DL of a linearly-precoded XL-MIMO system. {Differently from other papers adopting AS strategy, a distributed BS signal processing architecture is considered and the AS procedures are characterized in terms of the exchanged information between the processing nodes. Furthermore, we extend part of the results of \cite{marinello2020} with the proposition of AS algorithms for XL-MIMO that use the short-term fading coefficients instead of the long-term ones. Additionally, we address the problem of joint AS and PA in XL-MIMO sub-arrays using a decentralized RA algorithm. The proposed RA algorithm uses the Sherman-Morrison-Woodbury (SMW) formula to perform optimal power allocation (OPA) and AS in a decentralized fashion.}

The BS is constituted by multiple non-overlapping subarrays with dedicated remote processing units (RPUs), which perform independently channel estimation, precoding calculation and RA, mainly AS and PA. Each subarray is equipped with a fixed number of antenna elements and RF transceivers. Using the ZF precoding, the optimization goal is to maximize the SE subjected to the constraints of subarrays connections and maximum transmitted power. 

The contribution of this work is fourfold.
\textbf{\textit{i}}) Description of a distributed transceiver design for XL-MIMO based on a subarray switching architecture;
\textbf{\textit{ii}}) proposition of a centralized procedure based on the evolutionary heuristic GA to perform joint AS and PA to maximize the {SE} with subarray connection and maximum transmitted power constraints;
\textbf{\textit{iii}})  proposition of a distributed version of the GA procedure for joint AS and PA which achieves performance tight to the centralized one but with low-size coordination data and less number of executed operations;
\textbf{\textit{iv}}) extensive analyses of the proposed procedures in terms of number of symbols for training, coordination data size and number of floating point operations per second (flops).

The numerical results corroborate the GA-based procedures in achieving high performance, {specifically} in crowded XL-MIMO applications. Additionally, the decentralized GA version offers a good trade-off between performance, number of operations and coordination data size, outperforming the centralized procedures by adopting proper settings.

The rest of the paper is organized as follows. In Section \ref{sec:model} is described the system model, including the distributed subarrays processing at the BS. Next, {in} Section \ref{sec:joint-optimization} {are described} the centralized and distributed GA-based optimization procedures for joint AS and PA in XL-MIMO systems, while Section \ref{sec:antennas-selection} discusses two feasible AS procedures adopted as a result of decoupling the joint AS and PA optimization problem. Section \ref{sec:complexity} examines the complexity of the proposed algorithms. Extensive numerical results are discussed in Section \ref{sec:numerical-results}.  Final comments and conclusions are provided in Section \ref{sec:conclusions}.

% Notation
\subsection{Notation}
Boldface small $\mathbf{a}$ and capital $\mathbf{A}$ letters represent respectively vectors and matrices. Capital calligraphic letters $\mathcal{A}$ represent finite sets, and $| \mathcal{A} |$ denotes the cardinality of the set $\mathcal{A}$. $\mathbf{I}_n$ denotes the identity matrix of size $n$. $\{\cdot\}^T$ and $\{\cdot\}^H$ denote respectively the transpose and the conjugate transpose operators. $\textrm{diag}(\cdot)$, $\textrm{tr}(\cdot)$ and $\textrm{det}(\cdot)$ denote respectively the diagonal matrix, trace and determinant operators. $\left\lceil\cdot\right\rceil$ denotes the greatest integer operator. {$\binom{n}{k}$ denotes the binomial coefficient.} $\mathcal{CN}( \mu, \sigma^2 )$ is a circularly symmetric complex Gaussian distribution with mean $\mu$ and variance $\sigma^2$. $\mathbb{E}[\cdot]$ denotes the expectation operator.

% System model
\section{System Model}\label{sec:model}
\begin{figure}[!t]
\centering
\includegraphics[width=0.6\textwidth]{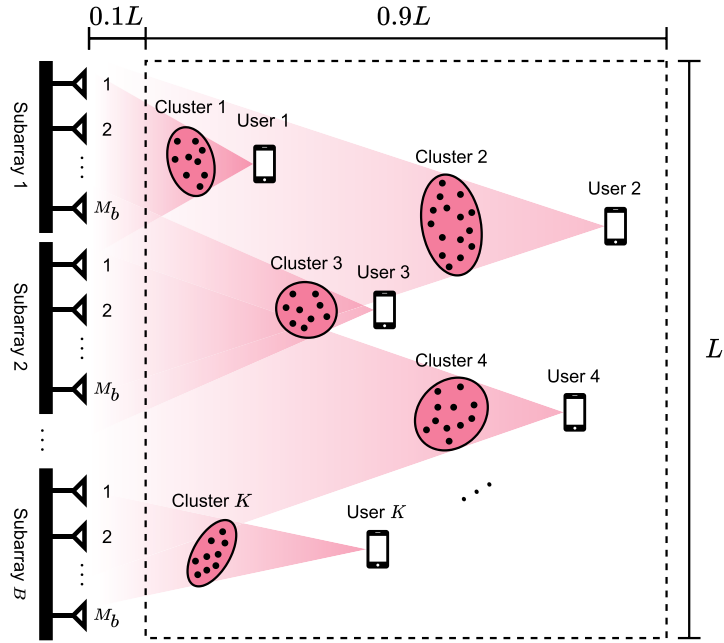}
\caption{{XL-MIMO system deployed inside a square cell with size $L$. The BS is a ULA with $M$ antennas divided into $B$ subarrays of $M_b$ antennas each one. The $K$ users are randomly distributed at a distance in the range $(0.1L,L)$ from the array.}}
\label{fig:cell-diagram}
\end{figure}

Consider the DL of a narrow-band multi-user XL-MIMO system with the BS equipped with $M$ antennas and $N$ RF transceivers serving $K$ single-antenna users, as is depicted in Fig. \ref{fig:cell-diagram}. During the DL, the BS uses $\eta_{\rm tr}$ symbols to perform channel estimation and $\eta_{\rm data}$ symbols to transmit the payload. We assume that the {time interval used to send} the total DL symbols $\eta_{\textsc{dl}} = \eta_{\rm tr} + \eta_{\rm data}$ is less than the channel coherence time.

\begin{figure}[!t]
\centering
    \includegraphics[width=0.6\textwidth]{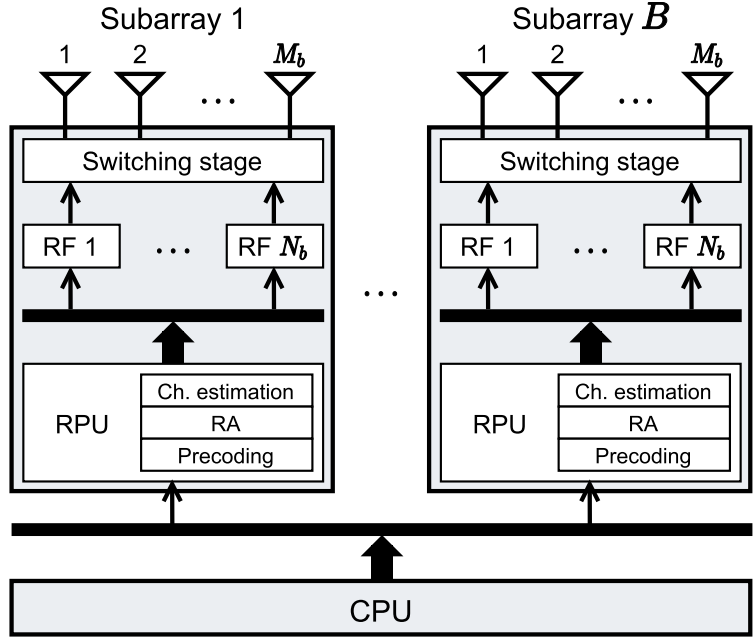}
    \caption{Diagram of the BS architecture for DL. The BS array is composed by $B$ subarrays containing $M_b$ antennas, $N_b$ RF transceivers and one RPU. Additionally, the BS has a CPU for subarrays coordination.}
	\label{fig:system-diagram}
\end{figure}

The array in the BS is composed of $B$ independent subarrays, each with $M_b$ antennas and $N_b < M_b$ RF transceivers. The subarrays are equipped with a RPU to perform, in a distributed way, {channel estimation}, precoding calculation and RA tasks, specially AS and PA procedures. In addition, the BS has a {central processing unit (CPU)} to coordinate the subarrays operation. Fig. \ref{fig:system-diagram} depicts all the described BS blocks.

\vspace{2mm}
\noindent\textit{Assumption 1 (Subarray switching stage):} A flexible switching stage is implemented in each XL subarray. This stage allows every antenna of the subarray $i$ to connect to any RF transceiver of it. Results in \cite{garcia-rodriguez2017} demonstrate that partially connected architectures introduce lower insertion loss than fully-flexible matrices, which allows the connection of any antenna in the entire array to any RF transceiver.

\vspace{2mm}

We assume that each subarray has perfect knowledge of the channel coefficients associated to its antennas. See \cite{sanchez2019} for details on channel acquisition in distributed signal processing architectures. {Besides, we deploy the ZF precoder to decode signals in each subarray. We adopt the technique in [21] to calculate the ZF precoder with low interconnection traffic, splitting the computations between the RPUs and the CPU.}

% Channel model
\subsection{Channel Model}
In the XL-MIMO scenario, spatial non-stationarities arise due to the large array physical dimensions and number of antenna elements. {Such} non-stationarities are addressed in the adopted channel model as the variation of the mean received power along the array, as in \cite{amiri2018, marinello2020}. The {path-loss}  coefficient associated to the BS antenna $m$ and the user $k$ is defined as
\begin{equation}
\label{eq:long-term-fading} % Long-term fading
\beta_{m,k}	= q_0d_{m,k}^{-\kappa}
\end{equation}
where $q_0$ is the path-loss attenuation at a reference distance, $d_{m,k}$ is the distance between the antenna $m$ and the user $k$ and $\kappa$ is the path-loss exponent.

Let {$\mathbf{R}_k \in \mathbb{C}^{M \times M}$,} $\mathbf{R}_k = \textrm{diag} ( [ \beta_{1,k} \; \cdots \; \beta_{M,k} ]^T )$ be the matrix with the long-term fading coefficients of the user $k$. The channel vector of the user $k$ is defined as
\begin{equation}
	\label{eq:channel-vector} % Channel vector
	\mathbf{h}_k
	= \mathbf{R}_k^\frac{1}{2}\mathbf{h}_k'
\end{equation}
where {$\mathbf{h}_k' \in \mathbb{C}^{M \times 1}$,} $\mathbf{h}_k' \sim \mathcal{CN} \left( \mathbf{0}, \mathbf{I}_M \right)$ is the short-term fading vector. From the users channel vectors, the channel matrix $\mathbf{H} \in \mathbb{C}^{M \times K}$ is defined as
\begin{equation}
	\label{eq:channel-matrix} % Channel matrix
	\mathbf{H}
	= \begin{bmatrix}
		\mathbf{h}_1 & \cdots & \mathbf{h}_K
	\end{bmatrix}
	= \begin{bmatrix}
		\mathbf{\underline{h}}_1^T & \cdots & \mathbf{\underline{h}}_M^T
	\end{bmatrix}^T
\end{equation}
considering $\mathbf{\underline{h}}_m \in \mathbb{C}^{1 \times K}$ as the channel vector with the coefficients associated to the antenna $m$.

During the DL, the BS activates a group of antennas represented by the set $\mathcal{S} \subseteq \left\{ 1,\dots,M \right\}$ such {that} $|\mathcal{S}| \leq N$. A partition of the set $\mathcal{S}$, {\it i.e.} $\{\mathcal{S}_b\}$, $\forall b = 1,\dots,B$, contains the index of the selected antennas in the subarray $b$. This set is defined such {that} $|\mathcal{S}_b| \leq N_b \; \forall b$, {meeting} the adopted subarray structure. The equivalent channel matrix of the active antennas is defined as a row-wise submatrix of $\mathbf{H}$, $\mathbf{H}_\mathcal{S} {\in \mathbb{C}^{|\mathcal{S}| \times K}}$. Similarly, the matrix $\mathbf{H}_{\mathcal{S}_b} { \in \mathbb{C}^{|\mathcal{S}_b| \times K}}$ contains only the channel vectors related to the active antennas in the subarray $b$.

Let $D_m \in \left\{ 0,1 \right\}, \; \forall m = 1,\dots,M$ be an indicator equal to 1 if the antenna $m$ is active during the DL and 0 otherwise. These indicators form the diagonal matrix $\mathbf{D} = \textrm{diag} ( [D_1 \; \cdots \; D_M ]^T )$. During the precoding and {SE} computations, it is required to calculate the matrix product $\mathbf{H}_\mathcal{S}^H\mathbf{H}_\mathcal{S}$ of the active antennas channel matrix. {Intended to} enable this computation by the distributed signal processing architecture, the Gramian matrix is defined as in the following.

\vspace{2mm}

\noindent\textit{Remark 1 (Gramian matrix):} Let $\mathbf{G}_m = \mathbf{\underline{h}}_m^H \mathbf{\underline{h}}_m, \; \forall m = 1,\dots,M$ be the Gramian matrix associated with the BS antenna $m$. The set $\mathcal{M}_b$ {is defined for} $b = 1,\dots,B$ as the group of antennas in the subarray $b$. The Gramian matrix associated to the $b$-th subarray includes only the active antennas inside it, and it can be written as
\begin{equation}
	\label{eq:gramian-subarray} % Subarray Gramian
	\mathbf{G}_{\mathcal{S}_b} = \mathbf{H}_{\mathcal{S}_b}^H\mathbf{H}_{\mathcal{S}_b} = \sum_{m \in \mathcal{M}_b} D_m \mathbf{G}_m
\end{equation}
Similarly, the array Gramian matrix considering only the active antennas is {defined as}
\begin{equation}
\label{eq:gramian-bs} % Array Gramian
\mathbf{G}_\mathcal{S} = \mathbf{H}_\mathcal{S}^H\mathbf{H}_\mathcal{S} = \sum_{m = 1}^M D_m \mathbf{G}_m
\end{equation}

{An upper bound for the system performance} considering the active antennas in the set $\mathcal{S}$, {namely the DL sum-capacity}, is calculated by \cite{gao2015}:
\begin{align}
\label{eq:capacity} % Capacity
C_\textsc{dpc}
&= \underset{\mathbf{P}}{\textrm{max}} \log_2 \textrm{det} \left( \mathbf{I}_K + \frac{1}{\sigma_z^2} \mathbf{PH}_\mathcal{S}^H\mathbf{H}_\mathcal{S} \right)\\
&= \underset{\mathbf{P}}{\textrm{max}} \log_2 \textrm{det} \left( \mathbf{I}_K + \frac{1}{\sigma_z^2} \mathbf{PG}_\mathcal{S} \right)\nonumber
\end{align}
where $\sigma_z^2$ is the additive noise power, while  {$\mathbf{P} = \textrm{diag} \left([ p_1 \; \cdots \;  p_K ]\right)$} denotes the matrix with the allocated power for each user. The powers $p_k, \; \forall k = 1,\dots,K$ are {defined} in order to meet the total power constraint $\sum_{k=1}^{K} p_k = P_{\max}$. The DL sum-capacity is achieved by the {\it dirty paper coding} (DPC) precoder, which has {prohibitive high-complexity for practical implementations}.

% Downlink signal
\subsection{Downlink Signal}
The data signal transmitted by the BS is defined as {$\mathbf{x} \in \mathbb{C}^{|\mathcal{S}| \times 1}$,}
\begin{equation}
	\label{eq:transmitted-signal} % Transmitted signal
	\mathbf{x}
    = \mathbf{FP}^{\frac{1}{2}}\mathbf{s}
\end{equation}
where $\mathbf{F} {\in \mathbb{C}^{|\mathcal{S}| \times K}}$ denotes the ZF precoding matrix, calculated by
\begin{align}
	\label{eq:precoding-matrix} % ZF precoding
	\mathbf{F}
	&= \mathbf{H}_\mathcal{S}\left( \mathbf{H}_\mathcal{S}^H\mathbf{H}_\mathcal{S} \right)^{-1}\\
	&= \mathbf{H}_\mathcal{S}\mathbf{G}_\mathcal{S}^{-1}\nonumber
\end{align}
$\mathbf{s} = [ s_1 \; \cdots \;  s_K ]^T$ denotes the vector of modulated data symbols such {that} $\mathbb{E} \left[ \|s_k\|_2^2 \right] = 1, \; \forall k = 1,\dots,K$ and $\mathbb{E} \left[ s_k^*s_{k'} \right] = 0, \; \forall k \neq k'$. The allocated powers in \eqref{eq:transmitted-signal} {are} calculated in order to {meet} the following power constraint
\begin{equation}
\label{eq:power-constraint} % Power constraint
\textrm{tr} \left[ \mathbf{P} \left( \mathbf{H}_\mathcal{S}^H\mathbf{H}_\mathcal{S} \right)^{-1} \right]
= \textrm{tr} \left( \mathbf{P} \mathbf{G}_\mathcal{S}^{-1} \right)
= P_{\max}
\end{equation}
Therefore, the entries of $\mathbf{P}$ depend on the active antennas set $\mathcal{S}$ and the PA policy.

The signal received by the users in the DL is defined as {$\mathbf{y} \in \mathbb{C}^{K \times 1}$,}
\begin{align}
	\label{eq:received-signal} % Received signal
	\mathbf{y}
	&= \mathbf{H}_\mathcal{S}^H \mathbf{F}\mathbf{P}^{\frac{1}{2}}\mathbf{s} + \mathbf{z}\\
	& = \mathbf{P}^{\frac{1}{2}}\mathbf{s} + \mathbf{z}\nonumber
\end{align}
where {$\mathbf{z} \in \mathbb{C}^{K \times 1}$,} $\mathbf{z} \sim \mathcal{CN} \left( \mathbf{0}, \sigma_z^2\mathbf{I}_K \right)$ denotes the additive noise vector.

Given the ZF precoding design, the {system SE} is calculated by
\begin{equation}
\label{eq:se} % ZF SE
{\textsc{SE}}
= \sum_{k=1}^K \log_2 \left( 1 + \frac{p_k}{\sigma_z^2} \right)
\end{equation}
which is equivalent to the {SE} of $K$ independent Gaussian channels with received signal-to-noise ratio (SNR) equal to $p_k/\sigma_z^2 \; \forall k$.

% Optimal power allocation policy
\subsection{Optimal Power Allocation (OPA) Policy}
The {OPA} policy is the one that solves the problem of maximizing the {system SE} at \eqref{eq:se}, subjected to the maximum power constraint {in} \eqref{eq:power-constraint}:
\begin{subequations}
\label{eq:opa-problem} % Optimal power allocation problem
\begin{align}
\underset{\mathbf{P}}{\textrm{maximize}} & \hspace{5mm}
{\textsc{SE}}
= \sum_{k=1}^K \log_2 \left( 1 + \frac{p_k}{\sigma_z^2} \right)\\
\textrm{subject to} & \hspace{5mm} \textrm{tr} \left[ \mathbf{P} (\mathbf{H}_\mathcal{S}^H\mathbf{H}_\mathcal{S})^{-1} \right] \leq P_{\max}\\
		& \hspace{5mm} p_k \geq 0, \; \forall k = 1,\dots,K
	\end{align}
\end{subequations}

The optimization problem in \eqref{eq:opa-problem} is equivalent to the well-known PA problem on independent Gaussian channels. It has an analytical closed-form solution derived by the Lagrange multipliers method (water filling solution). The optimal power distribution is calculated by \cite{he2013}:
\begin{equation}
	\label{eq:opa} % Optimal power allocation
	p_k
	= \left( \mu \left[ ( \mathbf{H}_\mathcal{S}^H\mathbf{H}_\mathcal{S} )^{-1} \right]_{k,k}^{-1} - \sigma_z^2 \right)^+
\end{equation}
where $( x )^+ = \textrm{max} ( x, 0 )$ and $\mu$ is a constant calculated by
\begin{equation}
	\label{eq:water-level}
	\mu
	= \frac{1}{K} \left\{ P_{\max} + \sigma_z^2 \textrm{tr} \left[ ( \mathbf{H}_\mathcal{S}^H\mathbf{H}_\mathcal{S} )^{-1} \right] \right\}
\end{equation}
If $p_k = 0$ for some user $k$, the PA problem including this user is not feasible. For this reason, the $k$-th user is deactivated and the power distribution is recalculated considering only the group of the remaining active users. This process {must} be repeated until a group of users which results in a feasible solution {is found}.

% GA-based antenna selection procedures
\section{Algorithm for Joint Antenna Selection and Power Allocation}\label{sec:joint-optimization}
The problem of jointly selecting the antenna-elements of the BS and allocating appropriate power amounts to maximizing the ZF {SE} given the constraints of maximum RF transceivers, subarray connections, and maximum power is formulated as
\begin{subequations}
	\label{eq:se-maximization} % SE maximization problem
	\begin{align}
        \underset{\mathbf{D},\mathbf{P}}{\textrm{maximize}} & \hspace{2mm}
	    \label{eq:se-maximization-objective} % Objective function
    {\textrm{SE}}
	    = \sum_{k=1}^K \log_2 \left( 1 + \frac{p_k}{\sigma_z^2} \right)\\
	    \label{eq:se-maximization-connections-constraint} % Connections constraint
	    \textrm{subject to} & \hspace{2mm} \sum_{m \in \mathcal{M}_b} D_m \leq N_b, \; \forall b \in \{1,\dots,B\}\\
	    \label{eq:se-maximization-power-constraint} % Power constraint
	    & \hspace{2mm} \textrm{tr} \left[ \mathbf{P} (\mathbf{H}^H\mathbf{DH})^{-1} \right] \leq P_{\max}\\
        \label{eq:se-maximization-binary-constraint} % Binary constraint
        & \hspace{2mm} D_m \in \{0,1\}, \; \forall m \in \{1,\dots,M\}\\
        \label{eq:se-maximization-positive-constraint} % Positive powers constraint
        & \hspace{2mm} p_k \geq 0, \; \forall k \in \{1,\dots,K\}
	\end{align}
\end{subequations}
{The objective function in \eqref{eq:se-maximization-objective} is the {system SE}. The constraints \eqref{eq:se-maximization-connections-constraint} are the subarray connections constraints, which allow the activation of a maximum of $N_b$ RF transceivers in each subarray. Also, the constraint \eqref{eq:se-maximization-power-constraint} ensures that the maximum transmitted power is equal to or less than $P_{\max}$. Moreover, the constraints \eqref{eq:se-maximization-binary-constraint} and  \eqref{eq:se-maximization-positive-constraint} define respectively the binary antenna association variables and non-negative allocated powers.}

Since $\mathbf{D}$ is binary constrained, the problem \eqref{eq:se-maximization} constitutes a non-convex combinatorial optimization problem. One approach to solve \eqref{eq:se-maximization} {comprises two steps:} {firstly, determining} the optimal active antennas set via exhaustive search assuming equal PA; after that, given the result $\mathbf{D}^\star$ from the exhaustive search, the allocated power matrix $\mathbf{P}^\star$ {is calculated adopting} the OPA policy in \eqref{eq:opa}.

The AS via exhaustive search considering the activation of all the RF transceivers requires testing $\binom{M_b}{N_b}^B$ candidate solutions, a number that attains prohibitive dimensions in the XL-MIMO {regime}. For instance, in a system with $B = 8$ subarrays equipped with $M_b = 64$ antennas and $N_b = 32$ RF transceivers, there is a number of feasible solutions {on} the order of $10^{146}$. Testing all these solution candidates in a timely manner is {impracticable}. An efficient alternative to the exhaustive search is to perform a guided search along the feasible set using an 
intelligent metaheuristic procedure. In this way, a good quality solution can be obtained in feasible time testing only a few candidates.

% Genetic algorithm
\subsection{Genetic Algorithm}\label{sec:genetic-algorithm}
One metaheuristic procedure adopted to solve many different combinatorial problems in wireless communications is the GA. This technique implements different search phases to  efficiently explore the feasible set and exploit the good candidates properties in order to find promising regions in the {feasible} sub-spaces. {Differently from exact optimization methods, evolutionary metaheuristics do not require convex objective functions or constraints.} In addition, the execution complexity can be fitted to the available computational burden by adjusting the input parameters and number of iterations. Despite the advantages, the GA, as well as other metaheuristics, does not ensure finding the optimal solution.

As the GA is a procedure inspired by {principles of genetics and natural selection}, it inherited several terms from {biology}. To simplify understanding, Table \ref{tab:genetic-algorithm-glossary} contains a glossary of some common GA terms adopted throughout this work. In the following, the implemented GA {procedures, phases and variables deployed} to solve the problem \eqref{eq:se-maximization} {are briefly} described.

\begin{table}[!t]
\centering
\caption{Glossary of the genetic algorithm terms}
%\resizebox{\columnwidth}{!}{
\label{tab:genetic-algorithm-glossary}
\begin{tabular}{|l|l|}
	\hline
	\textbf{Parameter} & \textbf{Description}\\
	\hline
Individual & Candidate solution for the optimization problem\\
	Population & Set of candidate solutions for the optimization problem\\
	Offspring & Set of candidate solutions generated during an iteration\\
	Gene & One optimization variable of the candidate solution\\
	Chromosome & Set of optimization variables of the candidate solution\\
	Generation & Genetic algorithm iteration\\
	Fitness & Objective function of the optimization problem\\
	Score & Value of the objective function for a candidate solution\\
	\hline
\end{tabular}
%}
\end{table}

\vspace{2mm}

\noindent\textbf{Optimization variables encoding:} The optimization variables of the problem \eqref{eq:se-maximization} are the antennas state indicators $D_m$ and the users allocated powers $p_k$. The powers $p_k$ are determined by the OPA, eq. \eqref{eq:opa}. Therefore, only the antennas indicators should be encoded as individuals. {Thus, $D_m$s are defined as genes} and the column vectors $[ \underline{\mathbf{d}}_{i, b} ]_m = D_m, \; \forall m \in \mathcal{M}_b, \; b = 1,\dots,B$ containing the optimization variables w.r.t. each subarray {represent} the chromosomes, where $i$ is the individual index. Every individual is {defined} by a vector {$\mathbf{d}_i \in \{0,1\}^{M \times 1}$,}
\begin{gather}
	\label{eq:individual} % Individual
	\mathbf{d}_i
	= \begin{bmatrix} \underline{\mathbf{d}}_{i, 1}^T & \cdots & \underline{\mathbf{d}}_{i, B}^T \end{bmatrix}^T
	= \begin{bmatrix} D_1 & \cdots & D_M \end{bmatrix}^T
\end{gather}

\noindent\textbf{Fitness function:} The fitness function considered for the implementation is the {ZF} SE defined in \eqref{eq:se}, with the power distribution computed by the OPA policy.

The implemented GA contains the following phases: \textit{a)} elitism, \textit{b)} tournament selection, \textit{c)} crossover and \textit{d)} mutation. These phases require the definition of the parameters: population size $N_p$, number of individuals for elitism $N_e$, number of tournaments $N_s$, crossover probability $p_c$ and mutation probability $p_m$. Each procedure is summarized in the sequel.

\vspace{2mm}

\noindent\textbf{Elitism:} The elitism aims to keep the best individuals of the current generation without change. At every generation, the $N_e$ best individuals are chosen as the first individuals of the next generation. Elitism ensures that {the SE obtained with the best AS indices of the GA iteration is always a non-decreasing value.}

\vspace{2mm}

\noindent\textbf{Tournament selection:} During the tournament selection, the individuals are pairwise randomly compared according to their score values. The winners of the $N_s$ tournaments become candidates for the crossover phase. {The selection step compares the sets of AS indices produced at each GA iteration according to the SE achieved by them.}

\vspace{2mm}

\noindent\textbf{Crossover:} The crossover phase aims to mix the chromosomes of the tournaments winners in order to obtain new solutions. {This phase exploits the good properties of the current set of AS indices}. Two tournament winners, named parent 1 and parent 2, are randomly selected to generate two new individuals. Each chromosome of child 1 has the probability $p_c$ of being inherited from parent 1 and $1 - p_c$ from parent 2. Considering child 2, every chromosome {has} the probability $p_c$ of being inherited from parent 2 and $1 - p_c$ from parent 1.

\vspace{2mm}

\noindent\textbf{Mutation:} The mutation phase aims to add random small changes at the offspring generated by crossover. This phase promotes the variability among the set of AS indices, exploring different regions of the feasible set. The chromosomes are mutated with probability $p_m$, {when} one random selected gene of the chromosome is flipped. To preserve the solutions' feasibility, the mutation phase is implemented by the scheme of Algorithm \ref{alg:mutation}. The set $\mathcal{P}_c$ denotes the offspring generated during the crossover, and $\mathcal{P}_m$ is the offspring after mutation.

\begin{algorithm}[!t]
\small
	\KwIn{Crossover offspring $\mathcal{P}_c$ $, p_m, B, M_b, N_b$}
	\KwOut{Mutated offspring $\mathcal{P}_m$}
    
	$\mathcal{P}_m \gets \emptyset$\;
    
	\For{$\mathbf{d}_i \in \mathcal{P}_c$}{
		\For{$b = 1:B$}{            
			\If{$\textrm{rand uniform} (0, 1) \leq p_m$}{
				$k \gets \textrm{rand discrete uniform}(1, M_b)$\;
				
				\If{$[ \mathbf{\underline{d}}_{i, b} ]_m == 0$ \hspace{-.8mm} and \hspace{-.8mm} $\sum_{j = 1}^{M_b} [ \mathbf{\underline{d}}_{i, b} ]_j == N_b$}{
					Go to line 5\;
				}
			
				$[ \mathbf{\underline{d}}_{i, b} ]_m \gets \textrm{flip}( [ \mathbf{\underline{d}}_{i, b} ]_m )$\;
			}
		}
		$\mathcal{P}_m \gets \mathcal{P}_m \cup \mathbf{d}_i$\;
	}
    \caption{Mutation procedure}
    \label{alg:mutation} 
\end{algorithm}

\vspace{2mm}

\noindent\textbf{Convergence:} There {are several mechanisms to check the GA convergence}. Herein, the implemented algorithm has two different criteria: the maximum number of generations $T_{\max}$ and the no improvement of the best {score} during the last $T_{\rm stall}$ generations.

Algorithm \ref{alg:genetic-algorithm} summarizes the implemented procedure, named {\it genetic algorithm for resource allocation} (GA-RA). The set $\mathcal{P}_0$ denotes the initial population, $\mathcal{P}_t$ the population of the generation $t$, $\mathcal{P}_s$ the winners of the tournament selection and $\mathcal{P}_{\rm temp}$ a temporary set for the elitism phase.

{\setstretch{1.0}
\begin{algorithm}[!t]
\small
\KwIn{$N_p, N_e, N_s, p_c, p_m, T_{\rm stall}, B, M_b, N_b, \mathbf{H}$}
	\KwOut{The best selected antennas set, $\mathbf{D}^\star$}
    
	$\mathcal{P}_0 \gets \emptyset$\;	
	$\mathcal{P}_0 \gets \mathcal{P}_0 \cup \textrm{N-AS}( \mathbf{H} )$ \textit{(Section \ref{sec:n-as})}\;
	
	\For{$i = 1:N_p - 1$}{
		$\mathcal{P}_0 \gets \mathcal{P}_0 \cup \textrm{rand individual}()$\;
	}
	
	\For{$t = 0:T_{\max}$}{
		$\mathcal{P}_{t+1}, \mathcal{P}_s, \mathcal{P}_c \gets \emptyset$\;
		$\mathcal{P}_{\rm temp} \gets \mathcal{P}_{t}$\;
		
		\For(\textit{Elitism}){$i = 1:N_e$}{
			$\mathbf{d}_e \gets \underset{\mathbf{d}_j}{\textrm{argmax}} \; \textrm{score}( \mathbf{d}_j ), \; \mathbf{d}_j \in \mathcal{P}_{\rm temp}$\;
			$\mathcal{P}_{t+1} \gets \mathcal{P}_{t+1} \cup \mathbf{d}_e$\;
			$\mathcal{P}_{\rm temp} \gets \mathcal{P}_{\rm temp} \backslash \mathbf{d}_e$\;
		}
		
		\For(\textit{Tournament selection}){$i = 1:N_s$}{
			$\mathbf{d}_{s_1},\mathbf{d}_{s_2} \gets \textrm{rand}( \mathcal{P}_t )$\;
			$\mathbf{d}_s \gets \underset{\mathbf{d}_j}{\textrm{argmax}} \left [ \textrm{score}( \mathbf{d}_{s_1} ), \;\textrm{score}( \mathbf{d}_{s_2} ) \right]$\;
			$\mathcal{P}_s \gets \mathcal{P}_s \cup \mathbf{d}_s$\;
		}
		
		\For(\textit{Crossover}){$i = 1:N_e$}{
			$\mathbf{d}_{c_1},\mathbf{d}_{c_2} \gets \textrm{rand}( \mathcal{P}_s )$\;
			$\mathbf{d}_{o_1},\mathbf{d}_{o_2} \gets \mathbf{0}_{M}$
			
			\For{$j = 1:B$}{
				\If{$\textrm{rand uniform}( 0,1 ) \leq p_c$}{
					$\mathbf{\underline{d}}_{o_1, j} \gets \mathbf{\underline{d}}_{c_1, j}$\;
					$\mathbf{\underline{d}}_{o_2, j} \gets \mathbf{\underline{d}}_{c_2, j}$\;
				}
				\Else{
				$\mathbf{\underline{d}}_{o_1, j} \gets \mathbf{\underline{d}}_{c_2, j}$\;
					$\mathbf{\underline{d}}_{o_2, j} \gets \mathbf{\underline{d}}_{c_1, j}$\;
				}
			}
			
			$\mathcal{P}_c \gets \mathcal{P}_c \cup \mathbf{d}_{o_1} \cup \mathbf{d}_{o_2}$\;
		}
		
		$\mathcal{P}_m \gets \textrm{mutation}( \mathcal{P}_c )$ \textit{(Algorithm \ref{alg:mutation})}\;
		
		$\mathcal{P}_{t+1} \gets \mathcal{P}_{t+1} \cup \mathcal{P}_m$\;
		
		$\mathbf{d}_{t+1}^\star \gets \underset{\mathbf{d}_i}{\textrm{argmax}} \; \textrm{score}( \mathbf{d}_i ), \; \mathbf{d}_i \in \mathcal{P}_{t+1}$\;
		
		\If(\textit{Stall convergence criterion}){$t > T_{\rm stall}$}{
			$\mathbf{d}_{\rm stall} \gets \underset{\mathbf{d}_i}{\textrm{argmax}} \; \textrm{score}( \mathbf{d}_i ), \; \mathbf{d}_i \in \mathcal{P}_{t-T_{\rm stall}}$\;
			\If{$\textrm{score}( \mathbf{d}_{t+1}^\star ) == \textrm{score}( \mathbf{d}_{\rm stall} )$}{
				Break the loop\;
			}
		}
	}
	
	$\mathbf{D}^\star \gets \textrm{diag}( \mathbf{d}_{t+1}^\star)$\;
	\Return{$\mathbf{D}^\star$\;}
    \caption{GA-RA}
    \label{alg:genetic-algorithm}
\end{algorithm}
}

% Distributed Genetic Algorithm
\subsection{Quasi-Distributed Genetic Algorithm}\label{sec:Q-dGA}
The proposed GA-RA procedure requires the entire channel matrix $\mathbf{H}$ knowledge at the CPU to compute the individuals score values. {Such} requirement is unfeasible in the XL-MIMO scenario due to the high bandwidth to transfer all the channel coefficients associated to {thousands} of antennas to the CPU. For this reason, one solution that does not depend on the knowledge of full CSI at the CPU is preferable.

One solution to avoid the {requirement of} full knowledge of {the} $\mathbf{H}$ {matrix} consists of performing local AS at each subarray, considering {fixed the AS indices in the other subarrays.} The contribution of these fixed AS indices can be calculated previously by the CPU and transmitted to the RPUs with reduced bandwidth and processing power resources. Therefore, each subarray can selects its antennas using the GA. The proposed {\it quasi-distributed genetic algorithm for resource allocation} (DGA-RA) implements this concept and is presented in the following.

Analyzing the fitness function of the GA-RA procedure in \eqref{eq:se}, one {can observe} that it depends on the inverse of the array Gramian matrix, $\mathbf{G}_\mathcal{S}^{-1} = ( \mathbf{H}_\mathcal{S}^H\mathbf{H}_\mathcal{S} )^{-1}$. The computation of $\mathbf{G}_\mathcal{S}^{-1}$ can be done from the subarrays Gramian matrices by
\begin{equation}
	\label{eq:inverse-gramian} % Inverse Gramian
	\mathbf{G}_\mathcal{S}^{-1} = \left( \sum_{b=1}^B \mathbf{G}_{\mathcal{S}_b} \right)^{-1}
\end{equation}
Therefore, the CPU can compute the inverse of the array Gramian matrix to calculate the GA-RA fitness function only with the subarrays Gramian matrices calculated locally at the RPUs. {Each subarray Gramian matrix has $K^2$ entries, while the channel matrix has $MK$. Therefore, calulating the contribution of the selected antennas at the CPU using the Gramian matrix strategy requires less bandwidth than by using the centralized strategy if $BK^2 < MK$ holds.}

Based on \eqref{eq:inverse-gramian}, the DGA-RA procedure operates as follows. Initially, each subarray selects an active antennas set based on a simple criterion, such as the {\it norm-based antenna selection} (N-AS) described in the subsection \ref{sec:n-as}. Then, the subarrays compute their Gramian matrices based on the selected set and transmit them to the CPU. At the CPU, the array Gramian matrix is computed by \eqref{eq:inverse-gramian} and transmitted back to the subarrays. Afterwards, every subarray performs local antenna selection by a GA implementation, considering that the other subarrays are fixed. To evaluate the fitness function in eq. \eqref{eq:se}, the subarrays compute the array Gramian inverse matrix adopting the SMW formula for matrix inversion, as follows.

\vspace{5mm}

\noindent\textit{Remark 2 (SMW formula):} The SMW formula \cite{golub2013} gives the inverse of the matrix $( \mathbf{A} + \mathbf{U}\mathbf{V}^H )$ from $\mathbf{A}^{-1}$, $\mathbf{U}$ and $\mathbf{V}$ by computing:% eq. \eqref{eq:smw-formula}.
\begin{equation}
\label{eq:smw-formula} % Sherman-Morrison-Woodbury formula
(\mathbf{A} + \mathbf{U}\mathbf{V}^H )^{-1}
		= \mathbf{A}^{-1} - \mathbf{A}^{-1} \mathbf{U} \left( \mathbf{I} + \mathbf{V}^H \mathbf{A}^{-1} \mathbf{U} \right)^{-1} \mathbf{V}^H \mathbf{A}^{-1}
\end{equation}
%\end{figure*}
Adopting this formulation, the array Gramian matrix can be calculated at the subarray $b$ during the iteration $n$ by letting
\begin{gather}
    \label{eq:smw-a} % SMW formula inverse of A
    \mathbf{A}^{-1}
    = \left( \mathbf{G}_{\mathcal{S}}^{(n-1)} \right)^{-1},\\
    \label{eq:smw-u} % SMW formula U
    \mathbf{U}
	=  \begin{bmatrix}
		- \left( \mathbf{H}_{\mathcal{S}_b}^{(n-1)} \right)^H & \left( \mathbf{H}_{\mathcal{S}_b}^{(n)} \right)^H
	\end{bmatrix},\\
	\label{eq:smw-v} % SMW formula V
\mathbf{V}^H = 	\begin{bmatrix}
\mathbf{H}_{\mathcal{S}_b}^{(n-1)}\\	\mathbf{H}_{\mathcal{S}_b}^{(n)}
\end{bmatrix},
\end{gather}
where the superscript $(n)$ denotes the variable during the $n$-th iteration of the DGA-RA procedure (proof in Appendix \ref{app:smw-formula}).

\begin{figure}[!b]
\centering
\includegraphics[width=0.6\textwidth]{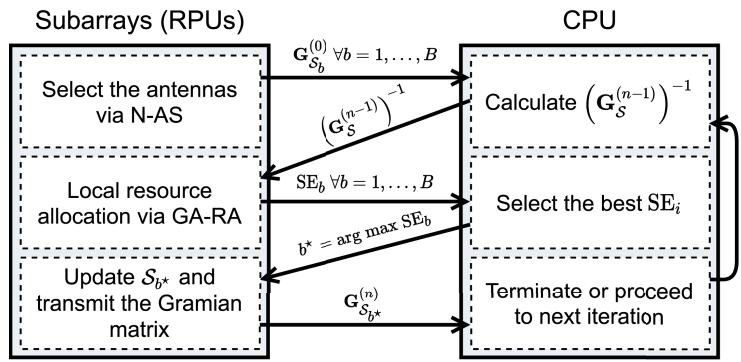}
\caption{Proposed DGA-RA procedure steps with coordination between the CPU and the RPUs. The superscript $(n)$ denotes the {$n$-th iteration.}} % variable value during the iteration $n$.}
\label{fig:dga-ra-procedure}
\end{figure}

After performing local AS, each subarray transmits their achieved SE values to the CPU. {The CPU updates the AS indices of the subarray that has achieved the maximum SE values at the iteration $n$}. Then, the CPU requests the subarray Gramian matrix {of} the updated subarray, and recalculates the inverse of the array Gramian matrix, $( \mathbf{G}_{\mathcal{S}}^{(n)} )^{-1}$. The process can be executed iteratively following the scheme depicted in Fig. \ref{fig:dga-ra-procedure}.

The GA implemented in the DGA-RA procedure is similar to that one described in the Algorithm \ref{alg:genetic-algorithm}, except for some details at the optimization variables encoding and the crossover phase. About the individual encoding, the optimization variables at each subarray are reduced from $M$ to $M_b$, since local AS is performed at each RPU. In addition, as the optimization variables consider only one subarray at each RPU, the individuals have two chromosomes: one represented by the first $M_b/2$ genes, and another composed by the remaining genes.

{Due to this new chromosome definition, one further procedure after the crossover phase is required to preserve the solution's feasibility.}
The chosen method is to deactivate antennas of individuals with more than $N_b$ antennas in a random fashion until they {become} feasible.

% Antenna selection procedures
\section{Antenna Selection Procedures}\label{sec:antennas-selection}
Two techniques to perform antenna selection are presented in the sequel, the DL {\it sum-capacity maximization antenna selection} ({SCMAX-AS}) and the N-AS method{, proposed respectively in \cite{gao2015}, \cite{garcia-rodriguez2017}}. The {goal of} solving only the antenna selection problem is to decouple the two RA problems associated to \eqref{eq:se-maximization} aiming at obtaining tractable formulations.

% Channel capacity maximization
\subsection{Antenna Selection for {DL Sum-Capacity} Maximization}\label{sec:cmax-as}
{Firstly, we analize equal power allocation {(EPA) strategy}, \textit{i.e.} $\mathbf{P} = \frac{P_{\max}}{K} \mathbf{I}_K$, intended to obtain a manageable optimization problem.} The problem of selecting the set of active antennas in order to maximize the {DL sum-capacity} with the constraints of maximum number of RF transceivers and subarray connections is formulated as {\cite{gao2015}}:
\begin{subequations}
\label{eq:capacity-maximization} % Capacity maximization problem
\begin{align}
\label{eq:capacity-maximization-objective} % Objective function
\underset{\mathbf{D}}{\textrm{maximize}} & \hspace{5mm}
{C_{\textsc{epa}}
= \log_2 \textrm{det} \left( \mathbf{I}_K + \frac{P_{\max}}{K \sigma_z^{2}} \mathbf{H}^H\mathbf{DH} \right)}\\
\textrm{subject to} & \hspace{5mm} \sum_{m \in \mathcal{M}_b} D_m \leq N_b, \; \forall b \in \{ 1,\dots,B \}\\
\label{eq:capacity-maximization-binary-constraint} % Binary constraint
& \hspace{5mm} D_m \in \{0,1\}, \; \forall m \in \{1,\dots,M\}
\end{align}
\end{subequations}
{Despite the concavity of the objective function in \eqref{eq:capacity-maximization-objective} \cite{dua2006}, the problem \eqref{eq:capacity-maximization} is not convex due to the binary constraint in \eqref{eq:capacity-maximization-binary-constraint}. Hence, we define a convex relaxation of \eqref{eq:capacity-maximization} by taking the variables $D_m$ in the range $(0, 1)$. This new problem, which can be solved with convex optimization tools, has the constraint \eqref{eq:capacity-maximization-binary-constraint} replaced by
\begin{equation}
	\label{eq:capacity-maximization-relaxed-constraint} % Relaxation of the binary constraint of the capacity maximization problem
	0 \leq D_m \leq 1, \; \forall m \in \{1,\dots,M\}
\end{equation}
Notice that the solution of the convex relaxation results in non-binary values for the active antenna indicators $D_m$, which is outside the original problem domain.}

One method for performing the antenna selection by solving the convex relaxation is to activate the $N_b$ antennas with the highest $D_m$ values at each subarray. This procedure is named in this work as {SCMAX-AS}, and is followed by the OPA policy in eq. \eqref{eq:opa}. This AS procedure gives near-optimal results, except for $N \ll M$ \cite{gao2015}. Therefore, in a XL-MIMO system where the number of available RF transceivers is much less than the array antennas, the achieved system SE with the {SCMAX-AS} algorithm will be sub-optimal.

% Norm-based
\subsection{Norm-Based Antenna Selection (N-AS)} \label{sec:n-as}
The N-AS procedure focus on selecting the subset of $N_b$ antennas with the highest channel vector norm values \cite{garcia-rodriguez2017}. {We adopt this method to initiate the population of the GA-based procedures due to its low computational cost.} The N-AS method solves the optimization problem formulated as 
\begin{subequations}
	\label{eq:norm-based} % Norm-based
	\begin{align}
		\underset{\mathbf{D}}{\textrm{maximize}} & \hspace{5mm}
		\Pi
		= \sum_{m=1}^M D_m \| \mathbf{\underline{h}}_m \|_2^2 \\
		\textrm{subject to} & \hspace{5mm} \sum_{m \in \mathcal{M}_b} D_m \leq N_b, \; \forall b \in \{ 1,\dots,B \}\\
		& \hspace{5mm} D_m \in \{0,1\}, \; \forall m \in \{1,\dots,M\}
	\end{align}
\end{subequations}
where the objective function consists of the sum of the {squared} norms of the channel vectors associated to the selected antennas.

The problem \eqref{eq:norm-based} can be solved quickly by selecting the $N_b$ antennas with the highest channel vector norms at each subarray. After selection, the PA is performed by the OPA policy in \eqref{eq:opa}.

% Complexity analysis
\section{Complexity Analysis}\label{sec:complexity}
The complexity of the presented procedures is evaluated in terms of the number of symbols required for channel acquisition, the size of the coordination data exchanged between the RPUs and the CPU, and the number of flops during execution.

% Training complexity
\subsection{Training}\label{sec:training}
In the following, we analyze the procedures in terms of training symbols for CSI acquisition. {The length of the mutually orthogonal pilot signals used to estimate the channel vectors at the BS depends on:} {\it a}) the number of users; {\it b}) the number of available RF transceivers; {\it c}) the number of antennas at the BS.

The number of symbols to acquire the entire channel matrix, required in all the presented procedures except {in} the N-AS, is $K \left\lceil\frac{M}{N}\right\rceil$. Particularly, the N-AS algorithm requires only the knowledge of the channel vector norms for selection. For this reason, the N-AS can be implemented without explicit channel estimation, supported by physical power-meters \cite{sanchez2019}. With this implementation, the N-AS requires a total of $2K$ symbols to operate. From this total, $K$ symbols are required to estimate the norms of the channel vectors, and the remaining $K$ symbols are used to estimate the channel vectors associated to the selected antennas.

% Coordination data size
\subsection{Coordination Data Size}\label{sec:exchanged-data}
The coordination data is defined as the data originated at the RPUs that is required at the CPU during the RA procedures. {Determining the coordination data size is crucial since it can grow tremendously in the XL-MIMO scenario.} In practical implementations, techniques as {\it data compression} helps alleviating the high interconnection bandwidth associated to the coordination data. However, such kind of consideration and optimization are out of the scope of this work.

Table \ref{tab:exchanged-data} contains the coordination data size associated to the considered RA procedures, detailing the type of required data in each one. The GA-RA and {SCMAX-AS} procedures require the entire channel matrix at the CPU, while the DGA-RA one relies on the subarrays Gramian matrices. On the other hand, the N-AS procedure does not require any CSI knowledge at the CPU for antenna selection purpose, being the most appealing technique in terms of the coordination data size.

\begin{table}[!t]
\centering
\caption{Coordination data exchanged between the RPUs and the CPU}
\label{tab:exchanged-data}
%\resizebox{\columnwidth}{!}{
\begin{tabular}{|l|l|l|l|}
	\hline
	\textbf{Procedure} & \textbf{Implementation} & \textbf{Data type} & \textbf{Data size}\\
	\hline
	GA-RA & Centralized & Channel matrix & $MK$\\
	\hline
	{SCMAX-AS} \cite{gao2015} & Centralized &  Channel matrix & $MK$\\
	\hline
	N-AS \cite{garcia-rodriguez2017} & Totally distributed & -- & --\\
	\hline
	DGA-RA & Quasi-distributed & Gramian matrix & $(B + N_{\rm it}) K^2$\\
	\hline
\end{tabular}
%}
\end{table}

\begin{figure*}[!b]
\centering
\subfigure[Varying $N_p$]{\includegraphics[width=.323\textwidth]{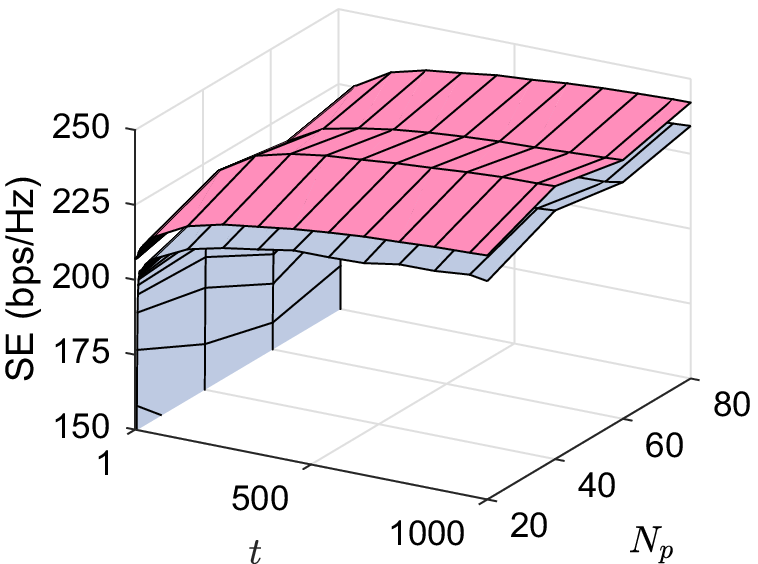}}
\subfigure[Varying $p_c$]{\includegraphics[width=.323\textwidth]{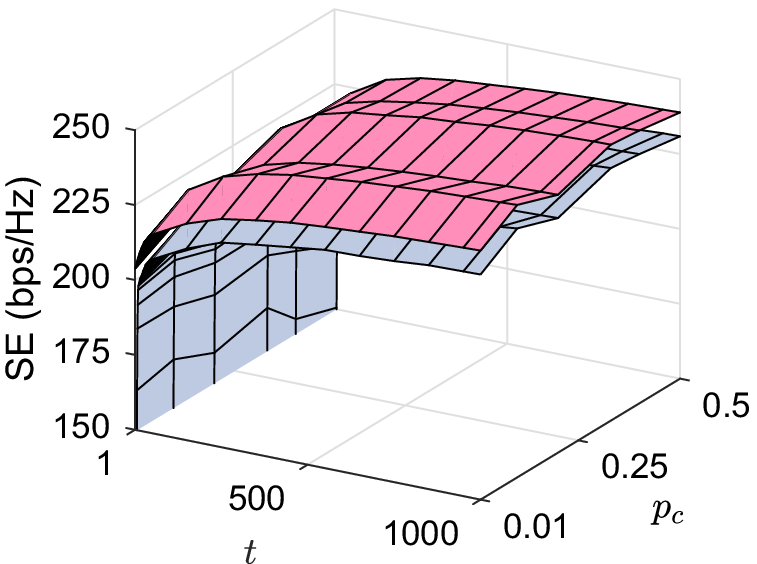}}
\subfigure[Varying $p_m$]{\includegraphics[width=.323\textwidth]{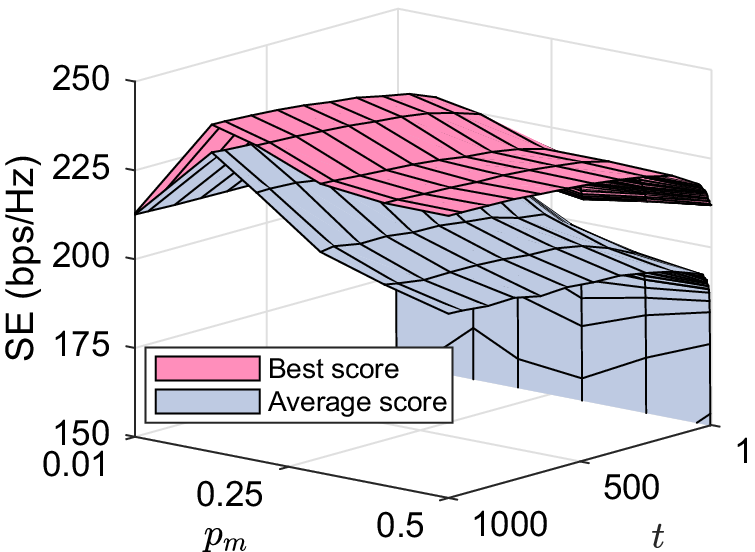}}
\caption{{Convergence of the GA-RA with the number of generations $t$ varying the GA input parameters $N_p$, $p_c$ and $p_m$. The "best" and "average" SE surfaces are obtained over 20 realizations. In each plot, the values of the remaining input parameters are given in Table \ref{tab:ga-parameters}.}}
\label{fig:ga-convergence} 
\end{figure*}

% Number of operations
\subsection{Number of Flops}\label{sec:operations}
The third complexity metric is the number of flops executed by each procedure. {The complexity analyses for the N-AS and the GA-based AS algorithms are as follows. The SCMAX-AS procedure is not considered due to the high complexity associated with computing the number of executed operations by the convex optimization solver.}

\vspace{2mm}

\noindent\textbf{N-AS:} The operations executed at each subarray on the N-AS procedure consists of calculating the channel vectors' norms then sorting the obtained values to get the $N_b$ largest ones. Assuming that the sorting operation has the complexity of the order $M_b \log ( M_b )$, the per-subarray flops for N-AS is
\begin{equation}
\label{eq:n-as-operations} % N-AS number of operations
\mathcal{C}_\textsc{n-as} = M_b ( 2K - 1 ) + M_b \log ( M_b )
\end{equation}

\noindent\textbf{GA-RA:} The complexity of the GA-RA method is dominated by the number of operations required for the evaluation of the GA fitness function, eq. \eqref{eq:se}. At the first iteration, the algorithm evaluate the fitness function for $N_p$ individuals. During the remaining iterations, $(T - 1)(N_p - N_e)$ fitness function evaluations are done, where $T$ denotes the total number of generations.

As the OPA policy involves simple computations, the complexity of the fitness function is reduced to the inversion of the array Gramian matrix. The flops to compute the array Gramian matrix inverse is derived in Appendix \ref{app:matrix-inversion-cholesky}. From this result, the total flops for the GA-RA algorithm is
\begin{equation}
\label{eq:ga-ra-operations} % GA-RA number of operations
\mathcal{C}_\textsc{ga-ra} = \left[ T (N_p - N_e) + N_e \right] \left( \frac{7}{3}K^3 + 2NK^2 - K^2 \right)
\end{equation}

\noindent\textbf{DGA-RA:} For the DGA-RA procedure, a similar approach to the one used for GA-RA can be followed. Despite that, the inverse of the array Gramian matrix is computed by the SMW formula, which is implemented with a different number of flops. The number of flops to obtain the inverse of the array Gramian matrix in the DGA-RA procedure is derived in Appendix \ref{app:matrix-inversion-smw}. Taking into account these differences and the fact that the DGA-RA procedure runs over $N_{\rm it}$ iterations, the total number of flops is given by:% the eq. \eqref{eq:dga-ra-operations}.
%\begin{figure*}
\begin{align}
\label{eq:dga-ra-operations} % DGA-RA number of operations
& \mathcal{C}_\textsc{dga-ra} = N_{\rm it} \left[ T (N_p - N_e) + N_e \right] \times \\
& \hspace{15mm} \times \left[ \frac{7}{3}N_b^3 + 2K^3 + N_b^2 ( 4K - 1 ) \right. + \notag\\
& \hspace{20mm} + K^2 ( 4N_b - 2 ) + N_b^2 ( 1 - 2K ) + K \bigg] \notag
\end{align}
%\end{figure*}

% Numerical results
\section{Numerical Results}\label{sec:numerical-results}
The numerical evaluations of the proposed methods as well as the benchmark techniques are presented in this section. The simulation system parameters are given in Table \ref{tab:simulation-parameters}. The users are randomly located inside a square cell of size $L$, and the BS is equipped with a uniform linear array (ULA) positioned on one side of the cell, as depicted in Fig. \ref{fig:cell-diagram}. Additionally, the users are random uniformly located at a distance in the range $(0.1L,L)$ from the array. {Although the results in the following are obtained for the ULA, they can be easily extended to other array form factors, such as the uniform planar one.}

\begin{table}[!t]
\centering
\caption{Simulation parameters}
\label{tab:simulation-parameters}
\begin{tabular}{|l|l|}
    \hline
    \textbf{Parameter} & \textbf{Value}\\
    \hline\hline
    Cell size & $L = 30$ m\\
    \# Users & $K \in [1,217]$\\
    Maximum transmitted power & $P_{\max} = 230$ $\mu$W\\
    Path-loss at the reference distance & $q_0 = -35.3$ dB\\
    Path-loss exponent & $\kappa = 3$\\
	Noise power & $\sigma_z^2 = -96$ dBm\\
	\hline
    \multicolumn{2}{c}{\textbf{\textit{Uniform Linear Array (ULA) Setup}}}\\
    \hline 
    \# Antennas & $M \in [32, 2048]$\\
    \# RF transceivers & $N \in [64, 256]$\\
    \# Subarrays & $B= \{2,4,8\}$\\
    \# Antennas per subarray & $M_b = M/B$\\
    \# RF transceivers per subarray & $N_b = N/B$\\ 
    \hline
\end{tabular}
\end{table}

Before comparing the proposed techniques, it is necessary to tune the GA-RA and DGA-RA GA input parameters in order to obtain a suitable performance-complexity tradeoff. The input parameter $N_p$, $p_c$ and $p_m$ values are selected using the iterated local search algorithm \cite{montero2014}. The number of individuals for elitism is equal to 10\% of the population size, and the number of tournaments is defined in order to fill the population after the elitism phase. Additionally, the stall convergence criterion parameter is approximately $30$\% of the maximum number of generations. The selected parameters for the GA-based procedures are listed in Table \ref{tab:ga-parameters}. Notice that the DGA-RA procedure is set to run 10 times less generations than the GA-RA, since the number of optimization variables decrease from $M$ at the GA-RA to $M_b$ in the DGA-RA procedure.

\begin{table}[!t]
\centering
\caption{Genetic algorithm parameters}
\label{tab:ga-parameters}
\begin{tabular}{|l|l|cc|}
	\hline
	\multirow{2}{*}{\textbf{Symbol}} & \multirow{2}{*}{\textbf{Description}} & \multicolumn{2}{c|}{\textbf{Parameter value}}\\
	&& {\bf GA-RA} & {\bf DGA-RA}\\
	\hline
	$N_p$ & Population size & 80 & 80\\
	$N_e$ & Elitism individuals & 8 & 8\\
	$N_s$ & Tournaments & 36 & 36\\
	$p_c$ & Crossover probability & 0.33 & 0.35\\
	$p_m$ & Mutation probability & 0.13 & 0.36\\
	$T_{\max}$ & Maximum generations & $10^3$ & $10^2$\\
	$T_{\rm stall}$ & Stall generations & 300 & 30\\
	\hline
\end{tabular}
\end{table}

{In Fig. \ref{fig:ga-convergence}, the quality of convergence of the GA-RA procedure is corroborated  varying the parameters $N_p$, $p_c$ and $p_m$ independently. Each surface is computed by averaging the achieved scores {over} 20 realizations. These results on the best and average SE scores among the generations $t$ confirm the parameters' values adopted in Table \ref{tab:ga-parameters}, while demonstrating a relative low tuning sensibility of the GA-RA convergence to the three input parameters.}

\begin{figure}[!t]
\centering
\includegraphics[width=0.6\textwidth]{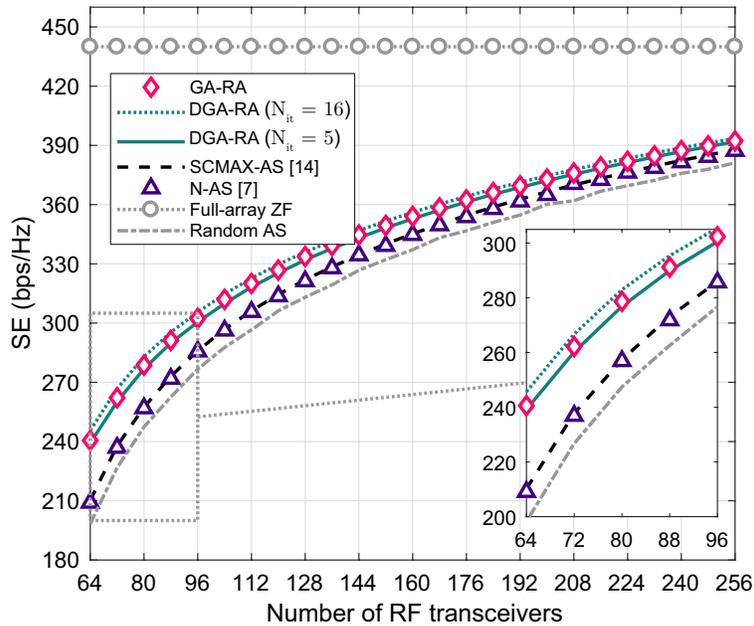}
\caption{{Comparison of SE {\it vs} the number of available RF transceivers}. $M = 512$, $B = 8$, $K = 50$ and, for the DGA-RA procedure $N_{\rm it} \in \{5,16\}$.}
\label{fig:se-rf-transceivers}
\end{figure}

Fig. \ref{fig:se-rf-transceivers} depicts the system SE {achieved by the proposed RA procedures versus the number of available RF transceivers. In addition to the proposed solutions, the SE attained by random AS scheme and using all the $M$ antennas are plotted as the lower and upper performance bounds, respectively.} The results consider $M = 512$, $B = 8$, $K = 50$ and $N_{\rm it}\in \{5,16\}$ for the DGA-RA procedure. Observing the Fig. \ref{fig:se-rf-transceivers}, one realize that the GA-based procedures achieve better {SE} results than the other ones. In the sequence, there are respectively the {SCMAX-AS} and N-AS. {As expected, all the performance curves are upper and lower bounded by the SE achieved using full-array ZF and random AS, respectively.} The {SE} gap between the procedures decreases as the number of RF transceivers increases. Analyzing the GA-based procedures, the DGA-RA achieves {SE} values tight to the GA-RA running with only five iterations. However, setting $N_{\rm it} = 16$ makes the DGA-RA system SE values outperform marginally the ones obtained by the GA-RA procedure. Therefore, the quasi-distributed procedure can achieve a performance comparable, or even better, to the fully centralized approach by adopting a sufficient number of iterations.

\begin{figure}[!t]
\centering
\includegraphics[width=0.6\textwidth]{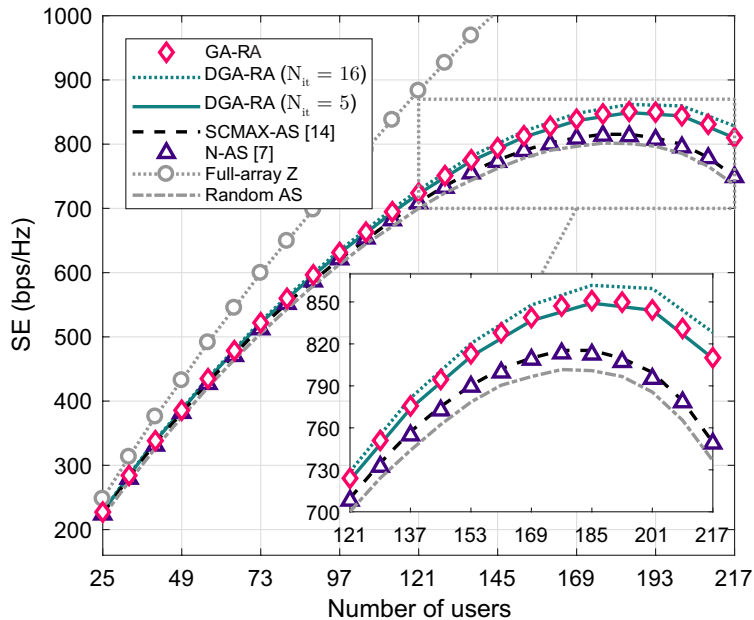}
\caption{{Comparison of SE {\it vs}} the number of users. $M = 512$, $B = 8$, $N = 256$ and, for the DGA-RA procedure $N_{\rm it} \in \{5,16\}$.
}
\label{fig:se-users} % SE against users figure
\end{figure}

In the following, Fig. \ref{fig:se-users} depicts the system SE achieved by the proposed RA procedures versus the number of users. These numerical results consider $M = 512$, $B = 8$, $N = 256$ and $N_{\rm it} \in \{5,16\}$ for the DGA-RA procedure. For better understanding, let $\mathcal{L} = K/N$ be the system {effective} loading factor. For all the proposed procedures, firstly the {SE} increases with $K$, assuming a decreasing behavior after a peak. This is due to the reduction of spatial degrees of freedom increasing the system loading factor, typically observed in linearly precoded systems \cite{marzetta2016}. Comparing the procedures, all of them get comparable {SE} values for a low loading factor. However, for high loading factor values, {typically $\mathcal{L} = 0.6$,} the GA-RA and DGA-RA procedures get {substantial} better results. Again, the DGA-RA outperforms the GA-RA in terms of {SE} by setting $N_{\rm it}= 16$. Combining the results in Figs. \ref{fig:se-rf-transceivers} and \ref{fig:se-users}, we conclude that the GA-based procedures perform with higher SE gains over the other {available AS schemes \cite{garcia-rodriguez2017, gao2015}} in crowded XL-MIMO scenarios, {\it i.e.}, when the loading factor is high, $\mathcal{L}>0.25$.

% Complexity numerical results
\subsection{Complexity Analysis}\label{sec:numerical-results-complexity}
\begin{figure}[!t]
{
\centering
\includegraphics[width=0.75\textwidth]{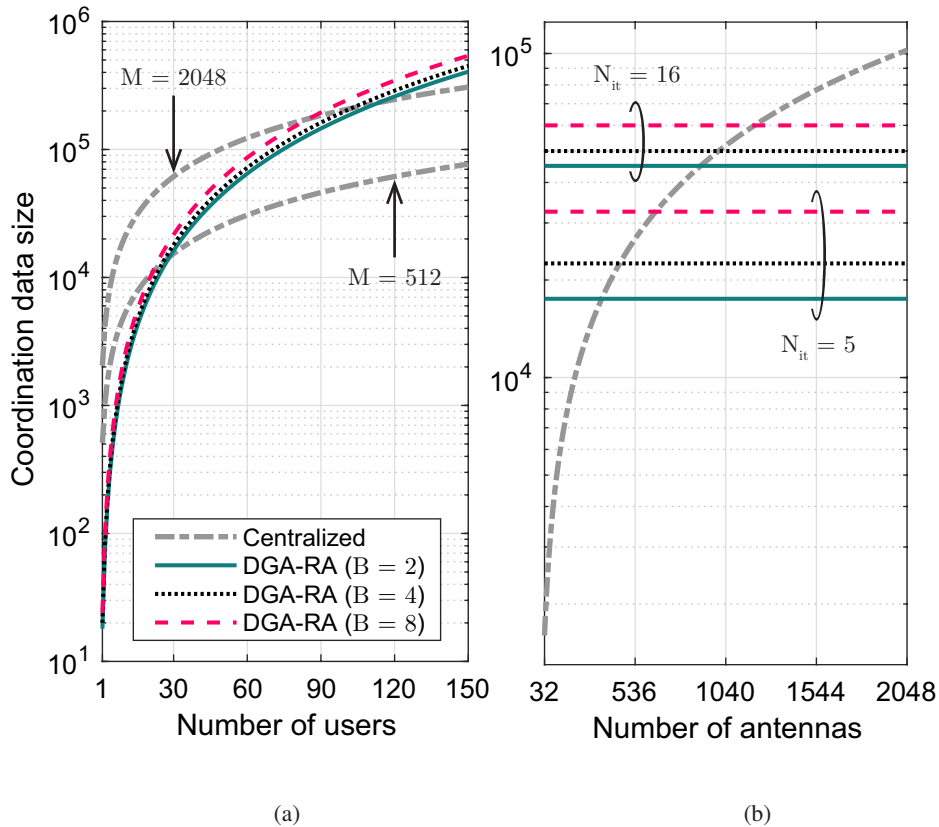}\\
}
\subfigure[]{\hspace{.7\textwidth}
\label{fig:coordination-data-users}
}
\subfigure[]{
\hspace{.001\textwidth}
\label{fig:coordination-data-antennas}
}
\vspace{-3mm}
\caption{{Coordination data size of the GA-based RA schemes {\it vs} the number of (a) users and (b) antennas. When it is not specified, $N_{\rm it} =  16$ and $K = 50$.}}
	\label{fig:coordination-data} % Coordination data size figures
\end{figure}

The numerical results in the following cover the computational complexity of the proposed procedures. In Fig. \ref{fig:coordination-data-users} the coordination data size {of the centralized procedures (GA-RA and SCMAX-AS) and the DGA-RA one versus} the number of users is illustrated. The curves are evaluated by the expressions in Table \ref{tab:exchanged-data}. The result considers $M \in \{512,2048\}$ and,for the DGA-RA procedure, $N_{\rm it} = 16$ and $B \in \{ 2,4,8 \}$. Comparing the RA approaches when the number of users is low, the quasi-distributed {one} get lower coordination data sizes than the centralized procedures. For higher {numbers of users}, the coordination data size associated to DGA-RA acquires larger values than the obtained by the centralized procedures. This point of inversion of behavior depends on the numbers of antennas, subarrays and iterations {w.r.t.} the DGA-RA procedure. It is worth mentioning that the coordination data size grows quadratically with $K$ for the DGA-RA procedure, while it grows linearly with $K$ for the centralized RA procedure.

Fig. \ref{fig:coordination-data-antennas} depicts the coordination data size {of the centralized procedures and the DGA-RA one} versus the number of antennas in the BS. The results consider $K = 50$ and, for the DGA-RA method, $N_{\rm it} \in \{5,16\}$ and $B \in \{ 2,4,8 \}$. The coordination data size grows linearly with $M$ in the centralized procedures, while for the DGA-RA procedure, it does not depend on $M$. In fact, this is the {primary aim} for choosing a distributed RA technique in XL-MIMO, in which the BS is equipped with an {asymptotically} high number of antennas.

\begin{figure}[!b]
{
\centering
\includegraphics[width=0.75\textwidth]{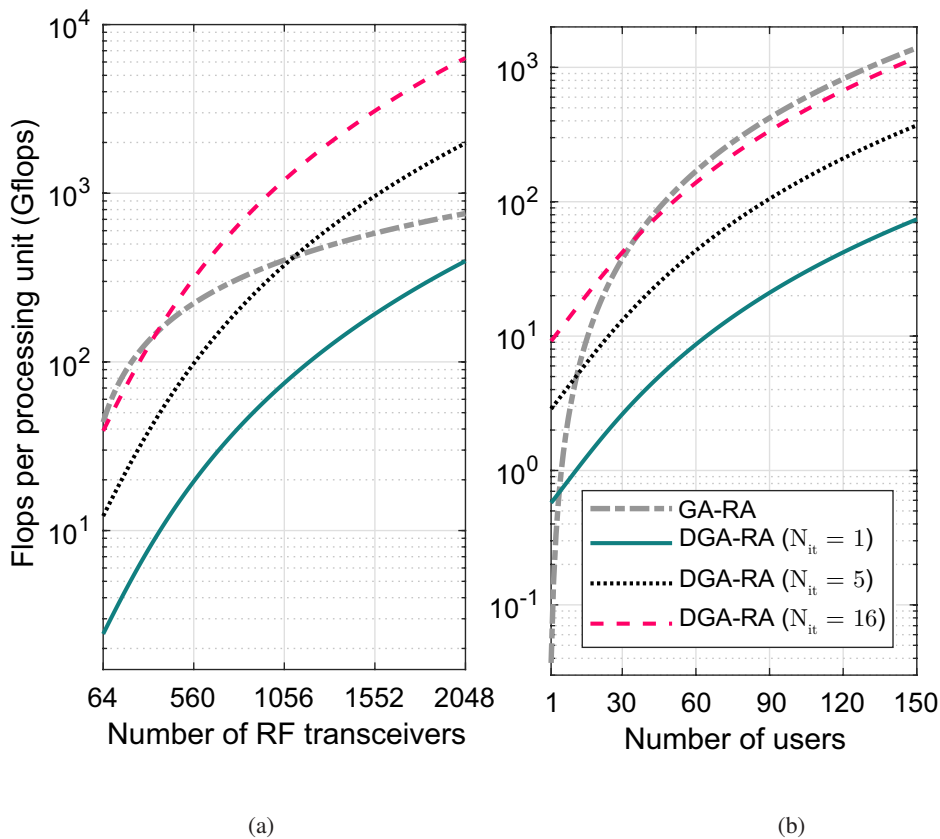}\\
}
\subfigure[]{
\hspace{.65\textwidth}
\label{fig:complexity-rf-transceivers}
}
\subfigure[]{
\hspace{.15\textwidth}
\label{fig:complexity-users}
}
\vspace{-3mm}
\caption{{Flops per processing unit of the proposed GA-based procedures versus the number of (a) available RF transceivers and (b) users. $B = 8$ and, when it is not specified, $K = 50$ and $N = 256$.}}	\label{fig:complexity} % Computational complexity figures
\end{figure}

{The} next results are related to the complexity in terms of flops. Fig. \ref{fig:complexity-rf-transceivers} {illustrates} the number of flops per processing unit {of the GA-based procedures} {versus} the number of available RF transceivers. The curves {are} evaluated by the eqs. \eqref{eq:ga-ra-operations} and \eqref{eq:dga-ra-operations}. Such results consider $K = 50$ and, {for the DGA-RA procedure,} $B = 8$ and $N_{\rm it} \in \{ 1,5,16 \}$. For low numbers of RF transceivers, the {flops'} values for the DGA-RA procedure are lower than the GA-RA algorithm. Again, after a point of inversion of behavior, the {flops'} values for GA-RA get lower than the ones for the quasi-distributed procedure. This point of changing of behavior decreases as $N_{\rm it}$ increases.

The {curves with the} number of flops per processing unit {of the GA-based procedures} {versus} the number of users are depicted in Fig. \ref{fig:complexity-users}. This result considers $N = 256$ and, {for the DGA-RA procedure,} $B = 8$ and $N_{\rm it} = \{1,5,16 \}$. For low numbers of users, the {flops'} values of the GA-RA procedure are lower than the ones get for the DGA-RA. However, this behavior inverts quickly, and the gap between the {flops'} values for both centralized and distributed procedures becomes constant. This constant behavior for large $K$ is due to the fact that both eqs. \eqref{eq:ga-ra-operations} and \eqref{eq:dga-ra-operations} grow asymptotically with $K^3$.

% Conclusions
\section{Conclusions}\label{sec:conclusions}
{This works proposes} a subarray switching architecture for the BS antenna array, while examining the problem of joint AS and PA optimization aiming at maximizing the {SE} of XL-MIMO systems with limited number of RF transceivers. Two GA-based {near-optimal and} low-complexity procedures {are} proposed. One is the centralized GA-RA, designed to operate with the entire channel matrix available at the CPU. The other is the quasi-distributed DGA-RA, based on the subarrays Gramian matrices. Both evolutionary metaheuristic optimization methods are {analysed} in terms of achieved SE, coordination data size and flops{, and compared} with {benchmarks, including two procedures from the literature, the SCMAX-AS and the N-AS followed by optimal PA}. Numerical results corroborate that the GA-based {AS and PA} procedures achieve high SE gains compared to the selected benchmarks, particularly in crowded XL-MIMO scenarios, {\it i.e.}, when the effective loading factor $\mathcal{L} > 0.25$. At the same time, the distributed DGA-RA method can outperform the other procedures with low-size coordination data and low computational complexity by taking the appropriate system operation settings.

% Acknowledgment
\section*{Acknowledgment}
This work was supported in part by the Coordenação de Aperfeiçoamento de Pessoal de Nível Superior - Brazil (CAPES) - Finance Code 001 (scholarship), in part by the Ministry of Science, Technology and Innovation (MCTIC), by the National Council for Scientific and Technological Development (CNPq) of Brazil under Grant  310681/2019-7, and in part by State University of Londrina -- Paraná State Government (UEL) and in part by the
Danish Council for Independent Research DFF-701700271.

% Appendices
\appendices

% Sherman-Morrison-Woodbury apprendix
\section{Local Computation of the Inverse of the Array Gramian Matrix via the Sherman-Morrison-Woodbury Formula}\label{app:smw-formula}
To compute the array Gramian matrix at the subarray $b$, the RPU {must follow these two steps. Firstly,} remove the contribution of the selected antennas at the subarray $b$ at the iteration $n-1$. {Then,} add the contribution of the selected antennas at the iteration $n$. Therefore, it needs to compute the inverse of the array Gramian matrix by the expression
\begin{equation}
\label{eq:subarray-gramian-inverse} % Inverse of the subarray Gramian
\left( \mathbf{G}_{\mathcal{S}}^{(n)} \right)^{-1}
	= \left( \mathbf{G}_{\mathcal{S}}^{(n-1)} - \mathbf{G}_{\mathcal{S}_b }^{(n-1)} + \mathbf{G}_{\mathcal{S}_b}^{(n)} \right)^{-1}
\end{equation}
which evaluation would be straightforward if all the terms were available at the subarray.

However, the subarray needs to compute $( \mathbf{G}_{\mathcal{S}}^{(n)} )^{-1}$ knowing only $( \mathbf{G}_{\mathcal{S}}^{(n-1)} )^{-1}$ and the local channel vectors, \textit{i.e.} $\mathbf{\underline{h}}_m \; \forall m \in \mathcal{M}_b$ for the subarray $b$. Writing the subarray Gramian matrices {of} \eqref{eq:subarray-gramian-inverse} in terms of the local channel matrices results in
\begin{align}
	& - \mathbf{G}_{\mathcal{S}_b}^{(n-1)} + \mathbf{G}_{\mathcal{S}_b}^{(n)}\nonumber\\
	& \hspace{10mm} = - \left( \mathbf{H}_{\mathcal{S}_b}^{(n-1)} \right)^H \mathbf{H}_{\mathcal{S}_b}^{(n-1)} + \left( \mathbf{H}_{\mathcal{S}_b}^{(n)} \right)^H \mathbf{H}_{\mathcal{S}_b}^{(n)}\nonumber\\
	\label{eq:subarray-gramian-prod-expansion} % U and V expressions
	& \hspace{10mm} = \begin{bmatrix}
		- \left( \mathbf{H}_{\mathcal{S}_b}^{(n-1)} \right)^H & \left( \mathbf{H}_{\mathcal{S}_b}^{(n)} \right)^H
	\end{bmatrix}
	\begin{bmatrix}
		\mathbf{H}_{\mathcal{S}_b}^{(n-1)}\\
		\mathbf{H}_{\mathcal{S}_b}^{(n)}
	\end{bmatrix}
\end{align}

From \eqref{eq:subarray-gramian-inverse} and \eqref{eq:subarray-gramian-prod-expansion}, it is possible to define the SMW formula variables, $\mathbf{A}^{-1}$, $\mathbf{U}$ and $\mathbf{V}^H$, in terms of the available information at the subarray as the eqs. \eqref{eq:smw-a}, \eqref{eq:smw-u} and \eqref{eq:smw-v}, respectively.

% Number of flops of the matrix inversion via Cholesky decomposition
\section{Flops to Compute the Inverse of the Array Gramian Matrix via the Cholesky Decomposition}\label{app:matrix-inversion-cholesky}
Initially, the computation of the array Gramian matrix is done by solving the product in \eqref{eq:gramian-bs}, which costs $2K^2N - K^2$ flops \cite{golub2013}. Afterwards, define the Cholesky decomposition of the array Gramian matrix as
\begin{equation}
	\mathbf{G}_\mathcal{S}
	= \mathbf{L} \mathbf{L}^H
\end{equation}
where $\mathbf{L}$ is a lower triangular matrix. The computation of $\mathbf{L}$ can be done with $K^3/3$ flops \cite{golub2013}. Then, each column of the inverse of the Gramian matrix can be computed solving the set of linear systems below by backforward substitution,
\begin{equation}
	\mathbf{L} \mathbf{L}^H \mathbf{x}
	= \mathbf{e}_i, \; \forall i = 1,\dots,K
\end{equation}
where $\mathbf{e}_i$ denotes the canonical basis vector, \textit{i.e.} a row vector with all entries equal to 0, except the entry $i$ which is equal to 1. Each linear system can be solved with $2K^2$ flops \cite{golub2013}, totaling $2K^3$ flops for all the columns of $\mathbf{G}_\mathcal{S}^{-1}$. Therefore, the total flops for the array Gramian matrix computation and inversion is equal to
\begin{equation}
	\label{eq:complexity-gramian-cholesky} % Complexity - Gramian matrix via Cholesky decomposition
	\mathcal{C}_{Chol.}
	= \frac{7}{3}K^3 + 2NK^2 - K^2
\end{equation}

% Number of flops of the matrix inversion via the Sherman-Morrison-Woodbury formula
\section{Flops to Compute the Inverse of the Array Gramian Matrix via the Sherman-Morrison-Woodbury Formula}\label{app:matrix-inversion-smw}
To count the flops to compute the matrix inversion by the SMW formula, the eq. \eqref{eq:smw-formula} is decomposed in six parts. The computations involved in each part and their respective flops are organized in Table \ref{tab:smw-complexity}. The flops in Table \ref{tab:smw-complexity} are counted assuming that the contribution of the selected antennas during the previous iteration is removed. {Such} assumption is reasonable since the expression in \eqref{eq:subarray-gramian-inverse} can be done sequentially, by keeping only the terms $- \mathbf{G}_{\mathcal{S}_b}^{(n-1)}$ or $\mathbf{G}_{\mathcal{S}_b}^{(n)}$ at a time.

All the parts include only simple matrix multiplications and sums, except for the part $\mathbf{Q}_3$. This part can be efficiently computed by the Cholesky decomposition approach followed by the backforward substitution procedure described in Appendix \ref{app:matrix-inversion-cholesky}. Therefore, the total flops required to compute the inverse of the array Gramian matrix via the SMW formula is equal to
\begin{align}
	\label{eq:complexity-gramian-smw} % Complexity - Gramian matrix via Sherman-Morrison-Woodbury formula
	&\mathcal{C}_{SMW}
	= \frac{7}{3}N_b^3 + 2K^3 + N_b^2 ( 4K - 1 )\\
	& \hspace{15mm} + K^2 ( 4N_b - 2 ) + N_b^2 ( 1 - 2K ) + K\nonumber
\end{align}

\begin{table}[!h]
\centering
\vspace{-2mm} \caption{Flops involved on the Sherman-Morrison-Woodbury formula computation}
\label{tab:smw-complexity}
\begin{tabular}{|l|l|l|}
	\hline
	\textbf{Symbol} & \textbf{Expression} & \textbf{Number of flops}\\
	\hline
	$\mathbf{Q}_1$ & $\mathbf{V}^H \mathbf{A}^{-1}$ & $2N_bK^2 - N_bK$\\
	$\mathbf{Q}_2$ & $\mathbf{I} + \mathbf{Q}_1 \mathbf{U}$ & $2N_b^2K - N_b^2 +N_b$\\
	$\mathbf{Q}_3$ & $\mathbf{Q}_2^{-1}$ & $7/3 N_b^3$\\
	$\mathbf{Q}_4$ & $\mathbf{U} \mathbf{Q}_3$ & $2N_b^2K - N_bK$\\
	$\mathbf{Q}_5$ & $\mathbf{I} - \mathbf{Q}_4 \mathbf{Q}_1$ & $2N_bK^2 - K^2 + K$\\
	$\mathbf{Q}_6$ & $\mathbf{A}^{-1} \mathbf{Q}_5$ & $2K^3 - K^2$\\
	\hline
\end{tabular}
\end{table}

\end{document}